# PARAMETRIC YIELD ANALYSIS OF MEMS VIA STATISTICAL METHODS

*Shyam Praveen Vudathu[1], Kishore Kumar Duganapalli[1], Rainer Laur[1]*
*Dorota Kubalińska[2], Angelika Bunse Gerstner[2]*

[1] Institute for Electromagnetic Theory and Microelectronics (ITEM)
[2] Center of Industrial Mathematics (ZeTeM)

University of Bremen
P.O. BOX 330440, 28334 Bremen, Germany
e-mail: svudathu@item.uni-bremen.de

**ABSTRACT**

This paper considers a developing theory on the effects of inevitable process variations during the fabrication of MEMS and other microsystems. The effects on the performance and design yield of the microsystems devices are analyzed and presented. A novel methodology in the design cycle of MEMS and other microsystems is briefly introduced. This paper describes the initial steps of this methodology that is aimed at counteracting the parametric variations in the product cycle of microsystems. It is based on a concept of worst-case analysis that has proven successful in the parent IC technology. Issues ranging from the level of abstraction of the microsystem models to the availability of such models are addressed.

## 1. INTRODUCTION

With the decreasing feature sizes and increasing demand for smaller devices, the future of Micro-Electro-Mechanical Systems (MEMS) and other microsystems is increasingly depending on the ability to design and fabricate devices with smaller feature sizes. Unfortunately, even the state-of-the-art process technologies have limitations while fabricating devices in the micro and nano scales. The limitations of the processes could be; inaccurate etching rate, mask misalignments, erroneous lithographic treatment, etc. These limitations in the processes could cause unexpected variations in the device dimensions like; variation in the line-width, beam-width, thickness of the membrane, etc. The variations in the design parameters are considered to be the process or statistical parameters of the design. Many efforts [7, 8] have been and are being done to accurately model the variations in the device geometry during the fabrication of MEMS structures. The modeling of processes parameters involves the consideration of different internal aspects. The process models describe the physical and geometric behavior of the process technologies.

The effect of process parameters on design yield of MEMS devices is analyzed. This is done by performing MEMS simulations that consider process models. The process models used in this work are Gaussian, Exponential or Uniform in nature. The parameters that are vulnerable to process variations are represented with their process models and the device is statistically simulated. A methodology that counteracts the effects of parametric variations on design yield of MEMS and other microsystems is briefly presented. This methodology dictates a change in the design parameter value by optimizing them in the design phase rather than the fabrication phase. It is based on the concept of worst-case analysis [4, 5] that has proven successful in the parent IC technology. This paper tests the possibility of the adoption of worst-case methods in the yield analysis of MEMS devices.

Section 2 starts with a demonstration of the necessity for statistical analysis of MEMS. Later, a brief review on the efforts being done in this domain is presented. Section 3 throws light on a tool developed in this work for performing sensitivity analysis of MEMS. In the same section, the simulation results of a capacitive pressure sensor are also presented. Section 4 briefly introduces the yield analysis techniques using worst-case methods and relates the results obtained in this work with these methodologies. This paper finally concludes by drawing conclusions and briefly describing the future work in section 5.

## 2. STATISTICAL ANALYSIS OF MEMS

### 2.1 Effect of Process Variations on MEMS

The necessity of statistical analysis of MEMS devices is being increasingly felt with the ever decreasing feature sizes. Many MEMS devices fulfill their specifications in the nominal sense (without the consideration of process variations). But the same designs when subjected to process variations in the manufacturing phase, partially fail to fulfill the specification values. This is explained with the help of a simple cantilever beam.

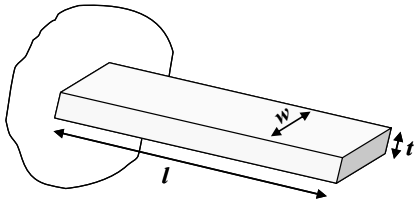

**Figure 1: A simple cantilever beam**

The simple beam spring formula of '*K*' is given by:

$$K = \frac{E \times t \times w^3}{l^3} \quad (1)$$

where  *t* is the thickness of the beam,
*w* is the width of the beam,
*l* is the length of the beam,
and  *E* is the Modulus of elasticity.

The resonant frequency '$f_r$' of the beam is extracted by changing slightly the mass at the free end of the beam. With '$f_0$' and '$f_1$' representing the initial and final resonant frequencies of the beam, the change in the mass can be represented as:

$$\Delta m = \frac{k}{4\pi^2}\left(\frac{1}{f_1^2} - \frac{1}{f_0^2}\right) \quad (2)$$

$$\Rightarrow \quad f_r \approx \sqrt{\frac{w^3}{l^3}} \quad (3)$$

The effect of process variations on this simple structure can be observed from the plot shown in Fig 2. The beam was nominally sized to better a specification value of 49KHz. With no process variations, the beam with a width of 2.0 microns, has a resonant frequency of 50KHz, which is greater than the specification value.
   A '1σ' variation in the width of the beam is considered to be reasonably in par with the usually observed variations at the fabsite. Fig. 2 shows a plot between the resonant frequency and the width of the cantilever beam with a '1σ' variation. The plot shows the dependence of the resonant frequency for two different values of the beam width.

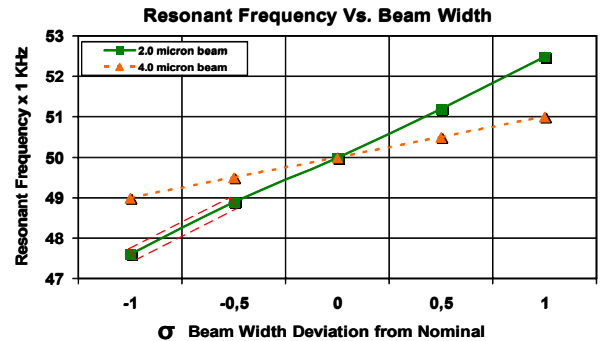

**Figure 2: Performance sensitivity to process variations in a cantilever beam**

From the plot, the nominal design that satisfies the resonant frequency specification is not doing the same in the entire range of the varied beamwidth. As the marked region of the line in Fig 2 indicates, a considerable amount of yield of this simple beam is affected due to process variations. Assuming a linear distribution, the yield of this design is roughly 78%. This affect of the process variation can be counteracted by increasing the beamwidth of the cantilever [3]. As the dotted curve (beamwidth = 4 microns) in the plot shows, the yield of the new design is close to 100%.

The yield of this design has been easily increased in this case because of:
- the presence of only two parameters
- considering only a single performance
- the availability of a simple and a good high abstraction model with a relatively simple structure.

Furthermore, not all devices can be blown up in size to have better yield. There are numerous devices that are ought to be small to serve a purpose (e.g. microfluidic channels, etc). All these reasons call for a better method where the device yield can be increased in a rather intelligent way.

### 2.2 Methods for Statistical Analysis of MEMS

This section presents a brief review on the existing statistical methods that could exploit process models in the design cycle of MEMS. The approach in [1] estimates the performance variations for general planar suspended MEMS structure for low frequency applications. In this work they develop the capabilities to find bounds on the performance parameters, given the bounds on the geometric variables. In [2], a comparison is drawn between various methods, and finally a mixture of methods like the Geometric Yield Optimization and the Taguchi method has been used in order to enhance the yield of an ADXL50 accelerometer. Nonetheless, Geometric yield optimization as implemented by the worst case distance (WCD) method, is employed only to bring the design closer to the target. This shows that the capabilities of the WCD methods have not been completely but only partly exploited in the design cycle of MEMS. The methodology considered in this paper can be generally applied to any MEMS device that satisfies a basic set of criterion. The chief criterion would be the level of abstraction of the model being used for the analysis. Though, principally model of any abstraction level could be analyzed, the usage of high level models is preferred because the simulation times of such models are comparatively less.

### 3. SENSITIVITY ANALYZER FOR MEMS

A sensitivity analysis can be performed on the MEMS design in order to identify the most relevant parameters, each of which influences the system performance largely. In sensitivity analysis, the system parameters are slightly perturbed and their influence on the performance is compared amongst different parameters. A Jacobian matrix is therefore a result of any first order sensitivity analysis. The Jacobian matrix when computed at the DC operation point of the device can also be used for the linearization of the performance around the DC operating point. This could later be used in the yield analysis of the design, which is about to be described in section 4.

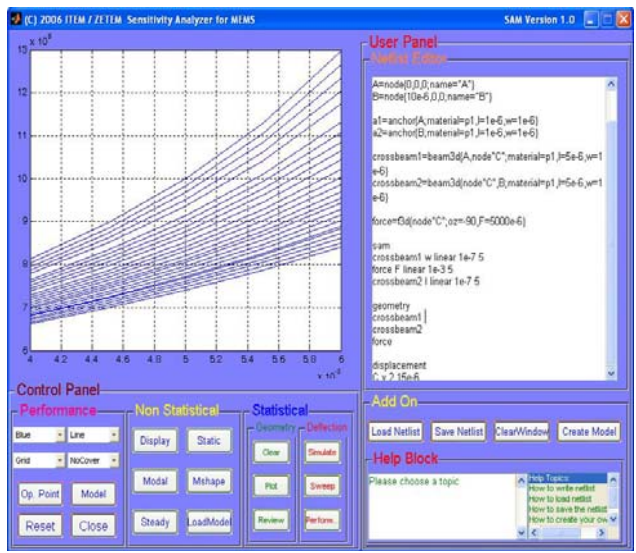

**Figure 3: Sensitivity Analyzer for MEMS - 'SAM'**

For performing this analysis, an interfacing tool 'SAM': Sensitivity Analyzer for MEMS was developed. The Sensitivity Analyzer for MEMS (SAM) developed in this work is based on the SUGAR simulator [9]. In order to perform a sensitivity analysis on a MEMS device, SAM induces process statistics into the design. This is done by appending an additional ".SAM" part to the conventional SUGAR netlist. The base of a "SAM" netlist is theoretically a SUGAR netlist, which is appended with more "SAM" commands to perform a statistical analysis. To begin with, some process models like the Gaussian, Exponential and the Uniform distribution models have been included. Nevertheless, other models could be easily integrated for analysis with this tool.

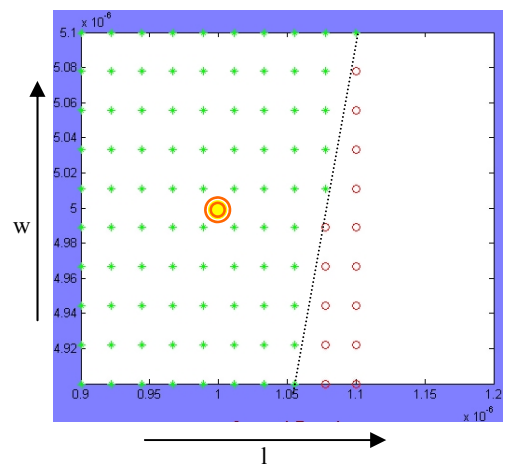

**Figure 4: Design space of a capacitive pressure sensor**

A simple capacitive pressure sensor was taken to be an example device in this work and has been simulated using SAM. The capacitive pressure sensor consists of a thin membrane and a chamber. More practical models of pressure sensors like that in [6] could be simulated using behavioral modeling languages like VHDL-AMS. On performing a sensitivity analysis, the most sensitive parameters were found to be the width and length of the membrane. Therefore, a yield plot between these two parameters in the design space is obtained by a two dimensional sweep. The force applied at the touchdown point (minimum force required for touchdown) was taken as the performance under consideration with a specification value of 5.5 micro Newtons. All the points with a '*' in the plot indicate that the point satisfied the specification value, while the points with 'o' indicate that the point doesn't satisfy the specification value. The yield of the design was observed to be **86%.** The same device could be analyzed with a different material for beam, which might result in totally new yield numbers.

## 4. STATISTICAL ANALYSIS BASED ON WORST-CASE METHODS

Amongst the various statistical methods, worst-case methods for yield analysis were highly successful in the analog and mixed signal domain of microelectronics and yielded excellent results. In this section, the considered yield analysis of MEMS using worst-case methods is briefly explained. Also, the relevance of the results obtained in section 3 to the worst-case methods is emphasized.

Using Monte Carlo analysis, the yield of the design can be calculated by performing huge number of simulations. The results of this huge number of simulations can be used in various ways. Yet there is no one single known way on how to improve yield of the design from Monte Carlo simulations. In worst-case analysis firstly, the performance of the design is linearized around the operating point. Secondly, the linearized specification boundary is used to calculate the worst-case distance from the nominal point. The advantage of the worst-case analysis lies in calculating the worst-case distance which can be directly equated to the yield of the design. Once the worst-case distance is known, larger the worst-case distance, larger is the yield of the design with respect to that particular specification.

With an assumption of a Gaussian distribution for the statistical parameters, the yield of a design '*Y*', could be directly calculated from the worst-case distance '*β*' using eq. (4).

$$Y \approx \frac{1}{2}\left(1 + erfc\left(\frac{\beta}{\sqrt{2}}\right)\right) \qquad (4)$$

The nominal point represented with a double lined dot in the Fig 4 represents the actual design point with zero variance. In Fig 4, the dotted line serves as discretion between the points that satisfy the specification and the points that don't. The dotted line hence is a specification boundary. The worst-case point is considered to be a point on the specification boundary and still the closest to the nominal point. Hence, the results obtained with SAM could be further extended to perform a yield analysis using worst-case methods.

## 5. CONCLUSIONS AND FUTURE WORK

The importance of the effect of process variations on MEMS devices has been demonstrated with an example of a cantilever beam. The necessity for new methods of yield analysis in the MEMS domain is becoming more obvious. SAM has been developed to meet the initial steps for yield analysis of MEMS devices using worst-case methods. SAM can be used to perform a sensitivity analysis on MEMS devices that are possible to be simulated with SUGAR.

An alternative approach would be to use other nodal analysis based simulators to simulate MEMS devices. This helps to implement the yield analysis techniques in a more generalized ambiance where most of the MEMS devices could be simulated. The implementation of the worst-case methods follows sensitivity analysis.